\documentclass[twocolumn]{aastex62}
\reportnum{Published in ApJ 2019, v872, id.76, 8pp.}
\shorttitle{Neutrino oscillations and decoherence}
\shortauthors{Penacchioni \& Civitarese}

\bibpunct{(}{)}{;}{a}{}{,}
\usepackage{savesym}
\savesymbol{iint}
\savesymbol{iiint}
\savesymbol{iiiint}
\savesymbol{idotsint}
\usepackage{amsmath}
\restoresymbol{AMS}{iint}
\restoresymbol{AMS}{iiint}
\restoresymbol{AMS}{iiiint}
\restoresymbol{AMS}{idotsint}
\usepackage{txfonts}
\usepackage{physics}
\usepackage{graphicx}			
\usepackage{natbib}	
\usepackage{xcolor}

\begin{document}

\correspondingauthor{A.V. Penacchioni}
\email{ana.penacchioni@fisica.unlp.edu.ar}

\author{A.V. Penacchioni}
\affil{IFLP (CONICET), La Plata, Argentina}

\author{O. Civitarese}
\affil{Department of Physics, University of La Plata (UNLP)\\
49 y 115 cc. 67, 1900 La Plata, Argentina}
\affil{IFLP (CONICET), La Plata, Argentina}

\title{Neutrino oscillations and decoherence in short-GRB progenitors}

\begin{abstract}
Neutrinos are produced in cosmic accelerators, like active galactic nuclei (AGNs), blazars, supernova (SN) remnants and gamma-ray bursts (GRBs). On their way to the Earth they experience flavor-oscillations. The interactions of the neutrinos coming from the source with other particles, e.g. intergalactic primordial neutrinos or heavy-mass right-handed neutrinos, in their way to the detector may transform the original wave packet in pointer states. This phenomenon, known as decoherence, becomes important in the reconstruction of processes at the source. In this work we study neutrino emission in short GRBs by adopting the Fireshell Model. We consider $e^{-}e^{+}$-pair annihilation as the main channel for neutrino production. We compare the properties of the neutrino-flux with the characteristic photon-signal produced once the transparency condition is reached. We study the effects of flavor-oscillations and decoherence as neutrinos travel from the region near the black-hole (BH) event-horizon outwards. We consider the source to be in thermal equilibrium, and calculate energy distribution functions for electrons and neutrinos. To compute the effects of decoherence we use a Gaussian model. In this scenario the emitted electron-neutrinos transform into pointer states consisting of $67.8\%$ electron-neutrinos and $32.2\%$ as a combination of mu and tau neutrinos. We found that decoherence plays an important role in the evolution of the neutrino wave packet, leading to the detected pointer states on Earth.
\end{abstract}


\section{Introduction}\label{introduction}

Short gamma-ray bursts (S-GRBs) are intense flashes of gamma-rays that last less than 2 seconds in the observer frame. It is widely accepted that S-GRBs originate from the merging of two compact objects, such as a neutron star (NS) and a BH, or two neutron stars (NS-NS). During the merging phase, angular momentum and energy losses are manifested as gravitational wave emission and electromagnetic radiation. In both cases the remnant is a BH of a few solar masses. There are different models which try to explain the observed emission of short GRBs; among others the Fireball Model \citep{1999PhR...314..575P} and the Fireshell Model \citep{2008AIPC..966...12B,2008mgm..conf.1989B,2008AIPC.1065..223B,2014styd.confE..73E}. 

The Fireball Model states that in the case of the NS-NS system an accretion-disk is formed around the newly born BH \citep{2014ARA&A..52...43B}. In the NS-BH case the same can occur if the NS is tidally disrupted outside the BH's event horizon. The rapidly rotating BH bends the magnetic-field-lines forming a double-jet perpendicular to the accretion-disk plane. A fraction of the electromagnetic radiation escapes in the form of gamma-rays, while another fraction goes into neutrino-antineutrino emission \citep{1992ApJ...395L..83N}. 

In the Fireshell Model scenario the NS-NS merging leads to a massive NS that exceeds its critical mass and gravitationally collapses to a BH with isotropic energy-emission of the order of $E_{iso} \gtrsim 10^{52}$ erg. Gravitational waves are produced \citep{2014ApJ...787..150O} together with GeV emission from the accretion onto the Kerr BH \citep{2018arXiv180207552R}. It has been shown \citep{2018ApJ...852..120B} that the accretion onto the NS generates neutrino-antineutrino emission in the case of long GRBs, and this emission has been explained as due to $e^{-}e^{+}$-pair annihilation. 

In this work we apply the Fireshell Model to explain neutrino emission in S-GRBs. We describe the conditions under which the neutrino emission takes place and we analyse the effects of flavor-oscillations and decoherence on neutrinos on their way from the source to the observer on Earth.

The work is organised as follows: in Section \ref{model} we describe the model.
In Section \ref{density} we derive the expressions for the electron and neutrino number densities, following a statistical treatment.
In Section \ref{flux} we compute the electron and neutrino fluxes at the source.
In Section \ref{oscillations vacuum} we analyze the effects of neutrino-flavor oscillations in vacuum, from the moment in which neutrinos are produced up to the time they reach the external crust.
In Section \ref{oscillations matter} we analyze the effects of neutrino-flavor oscillations in matter, as they propagate through the crust and interact with baryons.
In Section \ref{decoherence} we introduce the mechanism of decoherence, since neutrinos which leave the crust and propagate through the Universe towards the observer, interact with background intergalactic particles. We calculate the detected flux on Earth and compare it with the flux at source. The results are presented and discussed in Section \ref{numerical results}. Finally, in Section \ref{conclusions} we draw our conclusions.

\section{The model}\label{model}
\begin{figure*}
\centering
\includegraphics[width=\hsize]{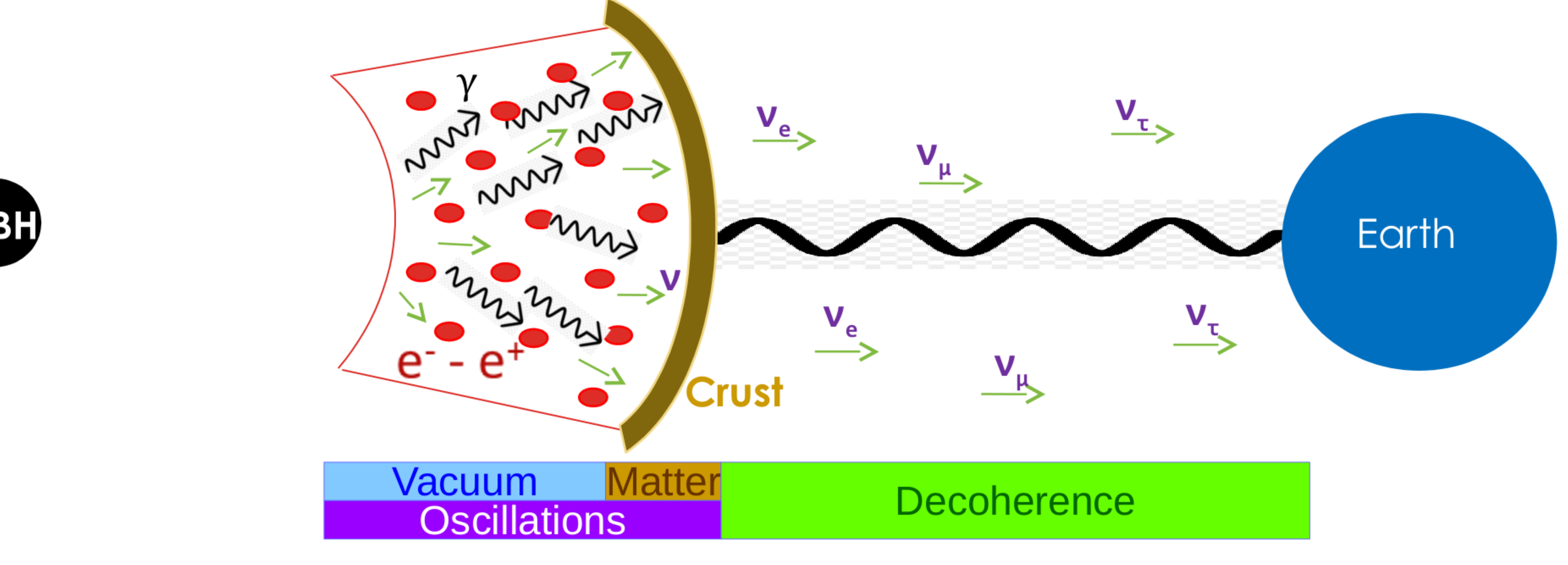}
\caption{Schematic representation (not to scale) of neutrino emission in the Fireshell Model. The two merging NS evolve into a BH. Vacuum polarization generates $e^{-}e^{+}$-pairs which annihilate to photons and neutrinos in the region delimited by the crust. Neutrinos travel towards the observer, oscillate in flavor and interact with intergalactic neutrinos.} 
\label{fig:diagrama}
\end{figure*}

A typical scenario for S-GRBs within the Fireshell Model is depicted in Figure \ref{fig:diagrama}. Two NS of masses $M_1$ and $M_2$, typically of the order of $1.6 - 2$ M$_\odot$, start spiraling together until they merge giving birth to a BH due to gravitational collapse. Let us consider for simplicity $M_1=M_2$. Only the core collapses, leaving a thin crust of a fraction of a solar mass. In the vacuum between the crust and the BH event-horizon a strong electric field is generated due to charge separation. When this field reaches the critical value, $E_c$, vacuum polarization takes place generating an $e^{-}e^{+}$-plasma. Typical densities for the electron-positron plasma are of the order of $10^{33}$ particles/cm$^3$. Some of these pairs annihilate giving neutrinos and antineutrinos which propagate outwards,  first in vacuum then through the crust formed by $e^{-}$, protons and neutrons, and finally through the intergalactic medium until they reach the observer on Earth \citep{2010RScI...81h1101H,2000AIPC..533..118S}. Another fraction of the $e^{-}e^{+}$-pairs produce thermal photons. Since at this stage the system is still opaque to radiation, the radiation pressure increases making the plasma expand until it reaches the crust. The whole system continues to expand until it reaches transparency. At this point the thermal photons escape. This is seen as a thermal spike in the spectrum called proper-GRB (P-GRB) \citep{2001ApJ...555L.117R}. The remaining material continues to expand while interacting with the circumburst medium producing the prompt emission.

Table \ref{tab:parameters} shows the values of the parameters of our model.

\begin{table} 
\centering
\caption{Parameters of our model.} 
\label{tab:parameters} 
\begin{tabular}{ccc}
\hline
Parameter& Symbol& Value\\
\hline
NS radius [km] & $R_{NS}$ & 10 \\
NS mass [$M_{\odot}$] & $M_{NS}$ & $ 2$\\
$e^{\pm}$ plasma density [part/cm$^3$] &$N_e^{\pm}$& $10^{33}$  \\
$e^{\pm}$ plasma temperature [MeV] & kT & 2.0\\
BH radius [cm] & $r_{BH}$ & $3.3 \times 10^5$\\
Crust internal radius [cm] & $r_{int}$ & $1.69 \times 10^8$\\
Crust external radius [cm] & $r_{ext}$ & $1.2 \times 10^{10}$\\
Source-detector distance [cm] & $D_L$ & $10^{28}$\\
Mass of the crust [$M_{\odot}$] & $M_{crust}$ & 0.1 \\
Density of the crust [g/cm$^3$] & $\rho_{crust}$ & $27.47$\\
Proton density in the crust [part/cm$^3$]& $N_p$& 0.25 $\rho_{crust}$\\
Neutron density in the crust [part/cm$^3$]& $N_n$& 0.25 $\rho_{crust}$\\
$e^-$ density in the crust [part/cm$^3$] &$N_e$& 0.50 $\rho_{crust}$\\
\hline
\end{tabular}
\end{table}

\section{Neutrino number density and energy}\label{density}

In order to calculate the number density and energy of the neutrinos created during the merging of the two NS, we follow a statistical treatment. We treat the neutrinos as a Fermi-Dirac gas in thermodynamical equilibrium at temperature $kT=2$ MeV \citep{1999A&A...350..334R}.
The neutrino emission zone is the same as the one occupied by the $e^{-}e^{+}$-plasma, a shell that extends from the BH event-horizon ($r_{BH}\approx 10^5$ cm) to the crust ($r_{crust}\approx 10^{10}$ cm).

The Fermi-Dirac distribution function for $T \neq 0$ is given by 
\begin{equation}\label{fermi}
f(E)=\frac{1}{e^{(E-\mu)/k_BT}+1},
\end{equation}
where $k_B$ is the Boltzmann constant and $E$ is the fermion energy. The parameter $\mu$ is known as the Fermi level. In the limit $T \rightarrow 0$, $f(E)$ becomes a step function $\theta (E -\mu)$: all the energy levels with $E < \mu$ are occupied, while all the others are empty.

\subsection{Electrons}\label{relativistic}

In the relativistic case the energy of the electrons is $E=\sqrt{p^2 c^2+m^2 c^4}$, the momentum in terms of the energy is given by $p= \frac{1}{c} \sqrt{E^2- m^2 c^4}$, and $dp=\frac{1}{c} \frac{E dE}{\sqrt{E^2- m^2 c^4}}$. The number of particles in the system is given by the expression

\begin{equation}\label{N}
N=\int f(E) d\Omega= \frac{4\pi V g_s}{(2\pi \hbar)^3}\int_0^\infty \frac{p^2 dp}{e^{(E-\mu_e)/k_BT}+1}.
\end{equation}
Here, $$d\Omega= \frac{d\overrightarrow{\mathbf{q}}d\overrightarrow{\mathbf{p}}}{(2\pi \hbar)^3} =  \frac{4 \pi V}{(2\pi \hbar)^3}p^2 dp$$ is the volume element in phase space and $g_s=2$ is the spin degeneracy factor.
Changing variables to $x=E-mc^2$ yields the number density $\rho_e=N/V$:
\begin{equation}\label{rhoe}
\rho_e=\frac{(2m)^{3/2}}{2 \pi^2 \hbar^3}\int_0^\infty \frac{x^{1/2} (1+x/2mc^2)^{1/2} (1+x/mc^2) dx}{1+e^{(x+mc^2-\mu_e)/kT}}.
\end{equation}

By making the substitutions $\eta_e=(\mu_e-mc^2)/kT$, $w=x/kT$ and $\beta=kT/mc^2$,  Eq.\ref{rhoe} becomes
\begin{equation}\label{rhoe_v2}
\rho_e=\frac{(2mkT)^{3/2}}{2 \pi^2 \hbar^3}[F_{1/2}(\eta_e)+\beta F_{3/2}(\eta_e)],
\end{equation}
where $$F_{r}=\int_0^\infty \frac{x^r (1+x\beta/2)^{1/2} dx}{1+e^{x-\eta_e}},$$ for $r=1/2,3/2,5/2$, etc.

A similar expression is obtained for the mean energy of the electrons as a function of temperature and density:
\begin{equation}
\frac{<E_e>}{\rho_e}=mc^2 \frac{F_{1/2}+2 \beta F_{3/2}+\beta^2 F_{5/2}}{F_{1/2}+\beta F_{3/2}}
\end{equation}
\subsection{Neutrinos}

Neutrinos have negligible masses compared to their energy ($E >> mc^2$), so $E \approx pc$. Following the same procedure as in Section \ref{relativistic} we find for the neutrino-number-density
\begin{equation}\label{rhonu}
\rho_{\nu}=\frac{N_{\nu}}{V}=\frac{4\pi g_s (k_BT)^3}{(2\pi \hbar c)^3}  F_2(\eta_{\nu}),
\end{equation}
where
$$F_{n}=\int_0^\infty \frac{x^n dx}{1+e^{x-\eta_{\nu}}},$$ for $n=1,2,3$,... and $\eta_{\nu} =\mu_{\nu}/kT$.

The neutrino mean energy is given by
\begin{equation}\label{E_fermidirac_ultrarel}
<E_{\nu}> = \frac{(k_BT)^4}{\pi^2 (\hbar c)^3} F_3(\eta_{\nu}),
\end{equation}
thus, the mean energy per neutrino is given by
\begin{equation}
\frac{<E_{\nu}>}{\rho_{\nu}}=kT \left( \frac{F_3(\eta_{\nu})}{F_2(\eta_{\nu})} \right).
\end{equation}

\section{Electron and neutrino fluxes at source}\label{flux}

With the parameters given in Table \ref{tab:parameters} and the formalism presented in Section \ref{density} we have performed a numerical search to determine the electron and neutrino chemical potentials $\mu_e$ and $\mu_{\nu}$, and with them the mean energies and spectral functions \citep{1968pss..book.....C}.

\begin{figure}
\centering
\includegraphics[width=\hsize]{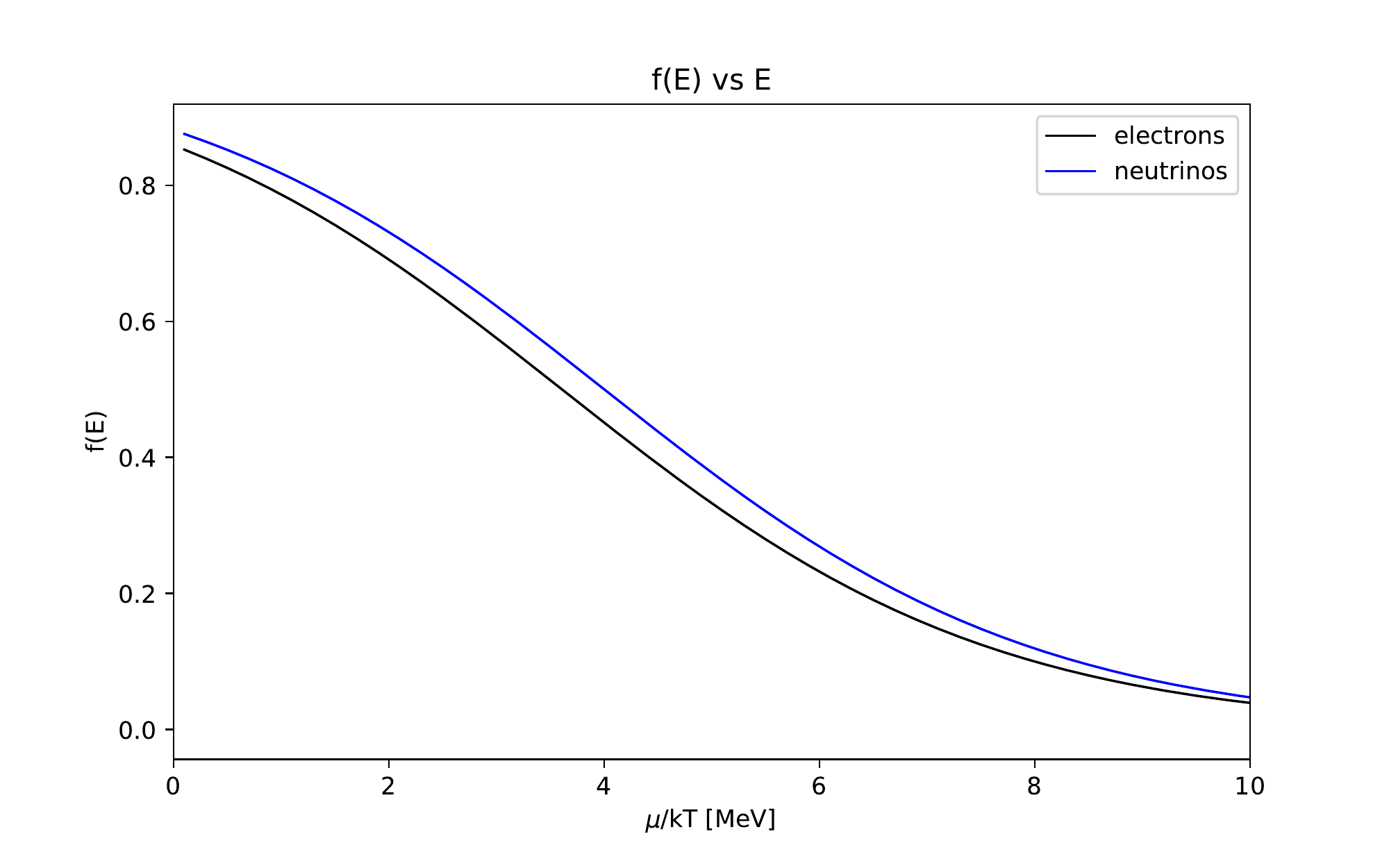}
\caption{Neutrino (upper curve) and electron (lower curve) occupation numbers for $kT=2$ MeV (see Eq.\ref{fermi}).} 
\label{fig:f_vs_E}
\end{figure}

The numerical search gives from Eqs. \ref{rhoe} and \ref{rhonu} the best values of $\eta_e$ and $\eta_{\nu}$ for a given density. The results are $\eta_e = 1.75$, $\eta_{\nu}= 2.0$, corresponding to $\mu_e =4.01$ MeV and $\mu_{\nu} =4.00$ MeV. Figure \ref{fig:f_vs_E} shows the occupation numbers $f(E)$ of Eq. \ref{fermi} for a plasma temperature of $2$ MeV \citep{1999A&A...350..334R}.

\begin{figure}
\centering
\includegraphics[width=\hsize]{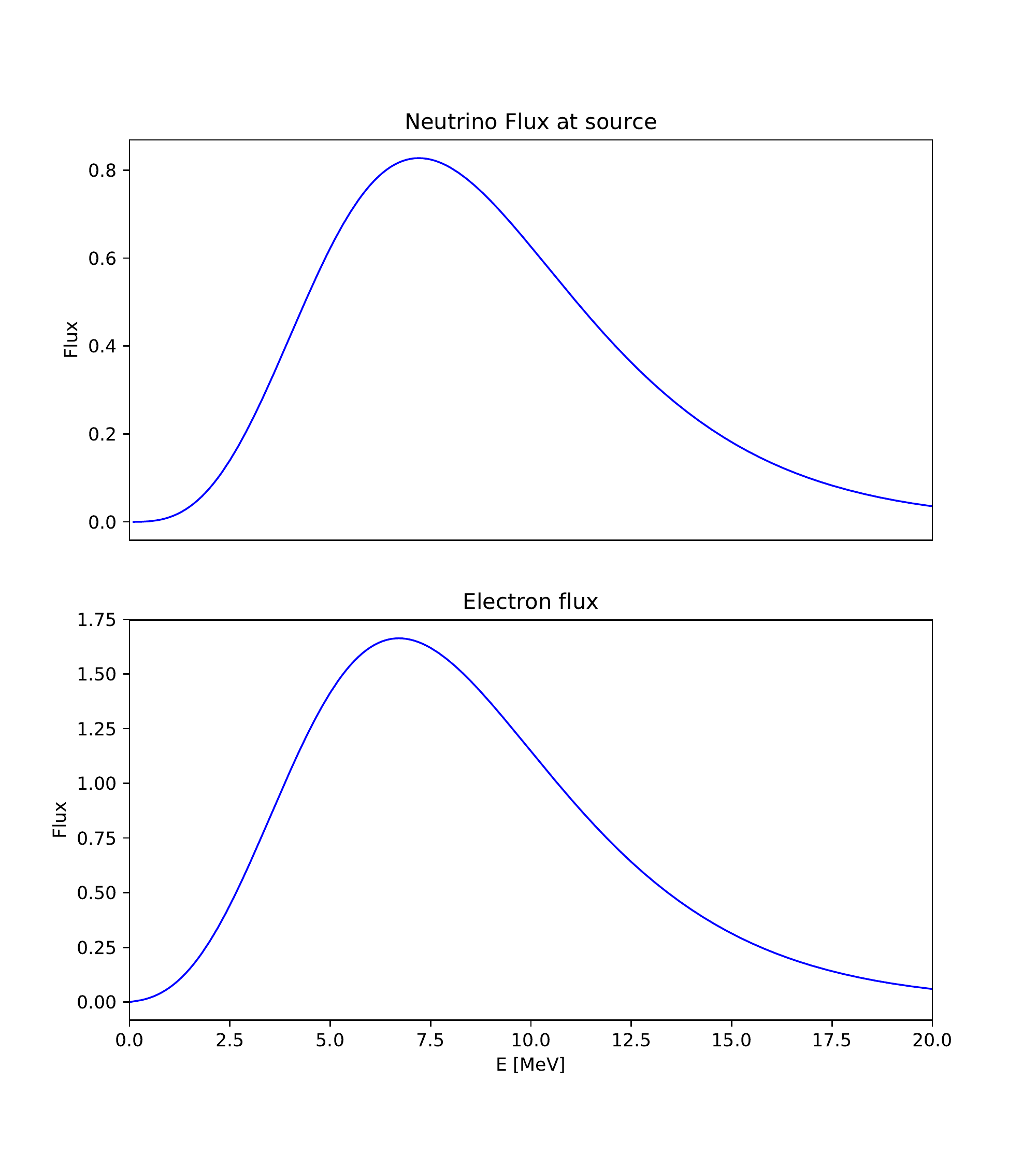}
\caption{Electron and neutrino fluxes, as a function of the energy, at the source (see Eq.\ref{eq:flux}).} 
\label{fig:flux}
\end{figure}

We have calculated the electron and neutrino fluxes inside the $e^{-}e^{+}$-plasma. Each flux is given by the ratio

\begin{equation}\label{eq:flux}
F_{e,\nu}=\frac{1}{\rho_{e,\nu}} \frac{d <E_{e,\nu}>}{dE}.
\end{equation}
Figure \ref{fig:flux} shows the results of Eq. \ref{eq:flux} for electron and neutrino fluxes in the region of the $e^{-}e^{+}$-plasma.

\section{Neutrino oscillations in vacuum}\label{oscillations vacuum}

As soon as they are created, neutrinos start to propagate outwards at nearly the speed of light from the region close to the event-horizon towards the crust. This region is opaque to radiation, but nothing prevents neutrinos from escaping. Because of the geometry of the source (see Figure \ref{fig:diagrama}) we shall consider propagation and oscillations in vacuum in the inner region between the BH and the crust.

Neutrinos oscillate because the flavor states in which they are created are a superposition of mass eigenstates. Since they have different masses they evolve with different phases. 

The Hamiltonian in the mass basis is given by 
\begin{equation}
H_m=
\begin{bmatrix}
    m_1 & 0 & 0\\
    0 & m_2 & 0\\
    0 & 0 & m_3\\
\end{bmatrix}
.
\end{equation}

 The mass hierarchies are denoted, as usual: $m_1 \leq m_2 << m_3$ (Normal Hierarchy), $m_1 << m_2 \leq m_3$ (Inverted Hierarchy), or $m_1 \approx m_2 \approx m_3$ (Degenerate Hierarchy). Adopting the normal hierarchy, and setting $m_1=0$, yields $m_2=0.00858$ eV and $m_3=0.0506$ eV. 
 
The neutrino mass Hamiltonian $H_m=diag(m_1,m_2,m_3)$ is transformed to the flavor basis by applying upon it the mixing matrix \citep{2016EPJC...76..339K,2000hep.ph....1311B}

\begin{equation}\label{Eq:U}
U=
\begin{bmatrix}
  c_{12} c_{13} & s_{12} c_{13} & s_{13} e^{-\delta} \\
  -s_{12} c_{23}-c_{12} s_{23} s_{13} e^{\delta} & c_{12} c_{23}-s_{12} s_{23} s_{13} e^{\delta} & s_{23} c_{13} \\
  s_{12} s_{23}-c_{12} c_{23} s_{13} e^{\delta} & -c_{12} s_{23}-s_{12} c_{23} s_{13} e^{\delta} & c_{23} c_{13} \\
\end{bmatrix}
,
\end{equation}
where $c_{ij}$ ($s_{ij}$) are the cosine (sine) of the mixing angles $\theta_{ij}$. Therefore,
\begin{equation}
H_f=U H_m U^{-1}.
\end{equation}
The parameters entering $U$ and $H_m$ are listed in Table \ref{tab:oscillation parameters1}. In the present analysis we have taken, for the Dirac CP-violating phase, the value $\delta/\pi=0$. Furthermore, and in order to separate the electronic flavor from linear combinations of the $\mu$ and $\tau$ ones, we apply to the flavor Hamiltonian the decoupling matrix \citep{2016EPJC...76..339K}

\begin{table}
\centering
\caption{Neutrino oscillation parameters, for the normal (NH) and inverted (IH) mass hierarchies. Solar ($\delta_{solar}^2$) and atmospheric ($\delta_{atm}^2$) squared-mass-differences and mixing angles ($\theta_{ij}$) are listed in the table. The values are taken from \citet{ParticleDataGroup}.}\label{tab:oscillation parameters1} 
\begin{tabular}{c}
\hline
$sin^2(\theta_{12})=0.297$\\
$sin^2(\theta_{13})= 0.0215$\\
$sin^2(\theta_{23})= 0.425$ (NH)\\
$sin^2(\theta_{23})= 0.589$ (IH)\\
$\delta_{atm}^2=m_3^2-m_1^2=2.56 \times 10^{-3}$ eV$^2$\\
$\delta_{solar}^2=m_2^2-m_1^2=7.37 \times 10^{-5}$ eV$^2$\\
\hline
\end{tabular}
\end{table}

\begin{equation}
D=
\begin{bmatrix}
  1 & 0 & 0 \\
  0 & 1/\sqrt{2} & 1/\sqrt{2} \\
  0 & -1/\sqrt{2} & 1/\sqrt{2} \\
\end{bmatrix}
.
\end{equation}

The `decoupled' flavor Hamiltonian reads 
\begin{equation}\label{Hdecoup}
H_f^{dec}= D H_f D^{-1},
\end{equation}
and from the diagonalization of this Hamiltonian we obtain the eigenvalues and eigenvectors for electron neutrinos $\nu_e$ and non-electronic neutrinos $\nu_x=\frac{1}{\sqrt{2}}(\nu_{\mu}\pm \nu_{\tau})$.

\section{Oscillations in matter}\label{oscillations matter}
Following the model sketch in Figure \ref{fig:diagrama}, neutrinos oscillate in vacuum until they reach the internal radius of the crust. At this point, they interact with matter. We assume the crust is formed by electrons, protons and neutrons in the amounts given in Table \ref{tab:parameters}. A matter Hamiltonian must be added to the flavor Hamiltonian in vacuum. For the matter Hamiltonian we consider a diagonal one 
\begin{equation}
H_{mat}= diag(V_m,0,0),
\end{equation}
where $V_m=\sqrt{2}G_F(N_e + N_p + N_n)$ is the matter potential, $G_F=8.963\times 10^{-44}$ MeV cm$^3$ is the Fermi constant and $N_e$, $N_p$ and $N_n$ are the electron, proton and neutron densities in the crust, respectively.

Because of the thickness of the crust ($\approx 10^2$ cm) the interactions with matter are negligible despite the values of the baryon densities. Therefore, we shall not take these interaction into account in our analysis.

\section{Decoherence}\label{decoherence}
Once the neutrinos arrive at the external radius of the crust they continue their way to the detector on Earth. Since the distance that they have to travel is of the order of $D_L \approx 10^{28}$ cm, corresponding to typical SGRB redshifts \citep{2016ApJ...831..178R}, decoherence effects \citep{2007dqct.book.....S} may take place due to interactions of the source neutrinos with neutrinos in the cosmic background. Decoherence effects are relevant in the reconstruction of the sequence of events starting from the primordial production of neutrinos and ending at their detection. What we would like to evaluate quantitatively is the difference between the composition of neutrinos of the source, as dictated by the neutrino-oscillation mechanism, and their time evolution governed by decoherence. The decoherence mechanism we have in mind in not kinematic and, as we said before, it is due to interactions with other particles like neutrinos which fill the space between the source and the detector. In order to achieve this goal we shall proceed to:
\begin{enumerate}
\item Calculate the density matrix from the diagonalization of the flavor Hamiltonian (Eq.\ref{Hdecoup}).
\item Construct the time evolution matrix which determines the time dependence of the density matrix.
\item Calculate the probability of detecting neutrinos of a given flavor on Earth.
\end{enumerate}

In what follows we present the corresponding theoretical details.

\subsection{Flavor eigenstates at $t=0$}

The density matrix for electron-neutrinos leaving the crust is
\begin{equation}\label{eq:rho0}
\rho_{\nu_e}= \ket{\phi_{\nu_e}} \otimes \bra{\phi_{\nu_e}}_{(t=0)}.
\end{equation}
With the amplitudes of the electron-neutrino eigenvalue obtained by the diagonalization of $H_f$, Eq.(\ref{eq:rho0}) is readily calculated. 

The density matrix (Eq.\ref{eq:rho0}) is that of a pure state, that is $\rho^2=\rho$, and its diagonalization yields the survival probabilities of the electron-, muon- and tau-neutrino channels, respectively.

\subsection{Time dependence of the density matrix}
To calculate the time dependence of the density matrix for neutrinos leaving the crust we add to the flavor Hamiltonian the interaction of the electron-neutrinos with the environment. For this, we follow the formalism presented in \citet{2017AIPC.1894b0006B} and \citet{2007dqct.book.....S}.
Accordingly, we construct the matrix
\begin{equation}
A=H_f^{dec}+diag(B \lambda_{coup} /2,0,-B \lambda_{coup} /2),
\end{equation}
where $\lambda_{coup}$ is the coupling constant and $B$ is a constant field acting on the neutrinos. The diagonalization of $A$ leads to the eigenvalues and eigenvectors needed to construct the evolution matrix $U(t)$ \citep{2007dqct.book.....S}, which is defined by the expression
\begin{equation}\label{U(t)}
U(t)= V \, \, diag(e^{i \Omega_n t}) \, \, V^{-1},
\end{equation}
where $V$ is the matrix of eigenvectors of $A$, $\Omega_n$, $n=1,2,3$, the associated eigenvalues, being both $V$ and $\Omega_n$ functions of the strength $B$. In writing Eq.(\ref{U(t)}) we use $\hbar=1$. In this picture the density matrix $\rho_{\nu_e}$ of Eq.(\ref{eq:rho0}) evolves with time as

\begin{equation}
\rho(t,B)= U(t) \, \, \rho_{\nu_e} \, \, U^{-1}(t). 
\end{equation}
If the strength $B$ is distributed like a Gaussian around $B=0$ with standard deviation $\sigma$, we integrate the matrix $\rho(t,B)$ in B so that its elements $[\rho(t)]_{ij}$ are: 
\begin{equation}\label{eq:rhoij}
[\rho(t)]_{ij}= \int{[\rho(t,B)]_{ij} \frac{e^{-B^2/2 \sigma^2}}{\sqrt{2 \pi}\sigma} dB}.
\end{equation}
A last diagonalization of $\rho(t)$ for $t$ sufficiently large, of the order of $L/c$, being $L$ the distance from the source to the detector and $c$ the speed of light, leads to the survival probabilities of neutrinos of a given flavor, in this case, of electron-neutrinos. These probabilities are needed in order to renormalise the neutrino flux at Earth, as explained below.

\subsection{Results for $\rho(t=0)$ and $\rho(t)$} \label{numerical results}

For the normal hierarchy and masses $m_1=0$, $m_2=0.00866$ eV, $m_3=0.0495$ eV and $\delta=0$ in Eq.(\ref{Eq:U}), we get
\begin{equation}
\rho_{\nu_e}(t=0)=
 \begin{bmatrix}
    0.680 & -0.124 & 0.449 \\
  -0.124 &  0.022 & -0.082 \\
   0.449 & -0.082 &  0.297 \\
 \end{bmatrix}
 .
 \end{equation}
The diagonalization of $\rho(t=0)$ gives 
\begin{equation}
P_{\nu_e \rightarrow \nu_e}(t=0)= 1,\\
P_{\nu_e \rightarrow \nu_{x}}(t=0)=0.
\end{equation}
To illustrate the effect of decoherence we calculate the time evolution given by Eq.\ref{eq:rhoij} with $\lambda_{coup}=1.0$ and $\sigma=20$.

For a sufficiently large number of oscillations in presence of the interactions due to the background and for the chosen parameterization, the density matrix is given by 
\begin{eqnarray}
\rho(t)& = & \rho_{(Re)}(t)+i\rho_{(Im)}(t) \nonumber \\
&=& 
\begin{bmatrix}
   0.680 & -0.0005 & 0.00002\\
 -0.0005 & 0.022 & -0.0002\\
 0.00002 & -0.0002& 0.296\\
\end{bmatrix} \nonumber \\
&+&i
\begin{bmatrix}
   0 & 0.005 &- 0.0008 \\
 - 0.005 & 0 & - 0.0003 \\
 0.0008 &  0.0003 & 0\\
\end{bmatrix}
,
\end{eqnarray}
which is no longer the density matrix of a pure state, since $\rho^2 \neq \rho$. 

\begin{figure}
\centering
\includegraphics[width=1.1\hsize]{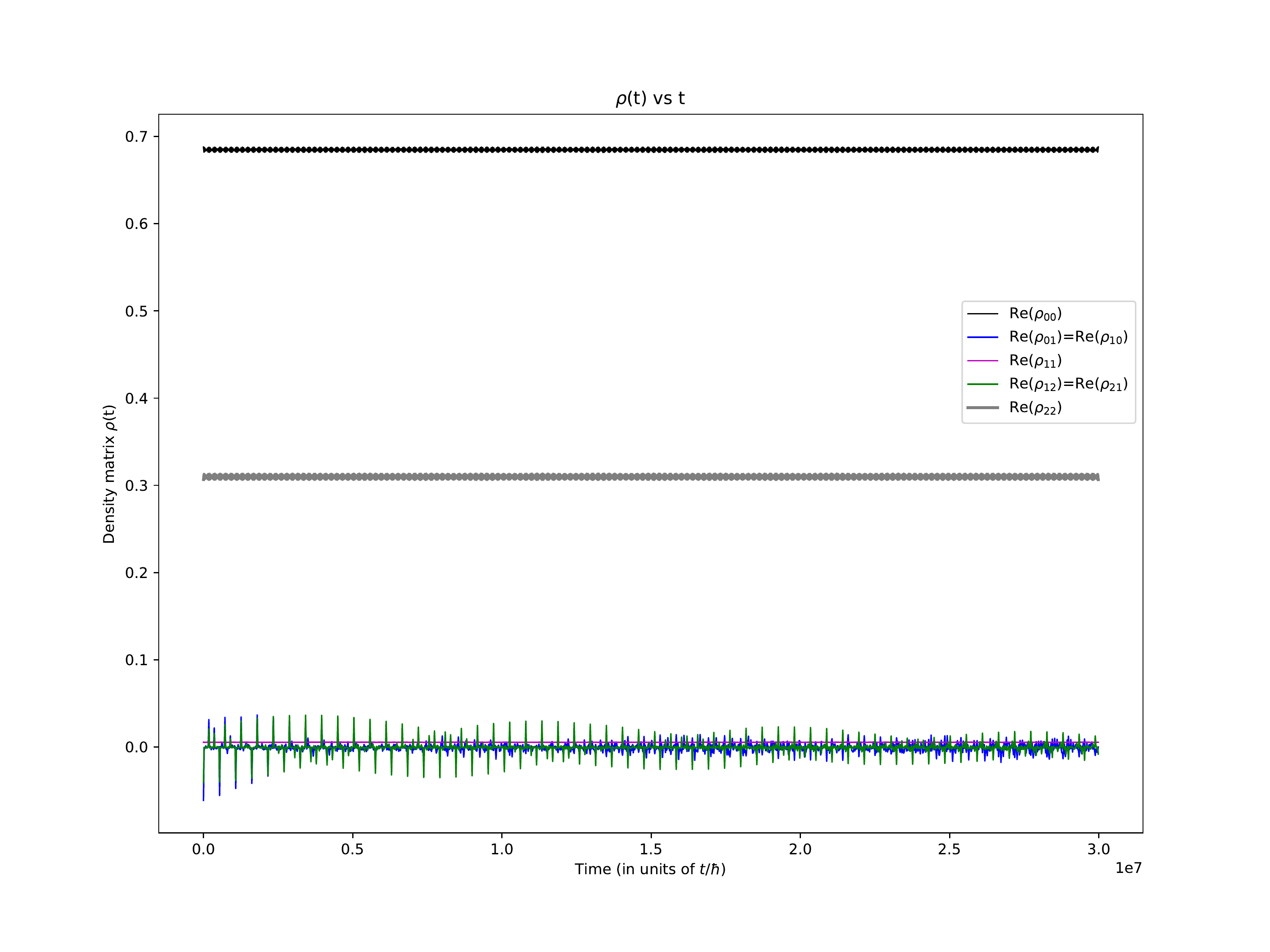}\\
\includegraphics[width=1.1\hsize]{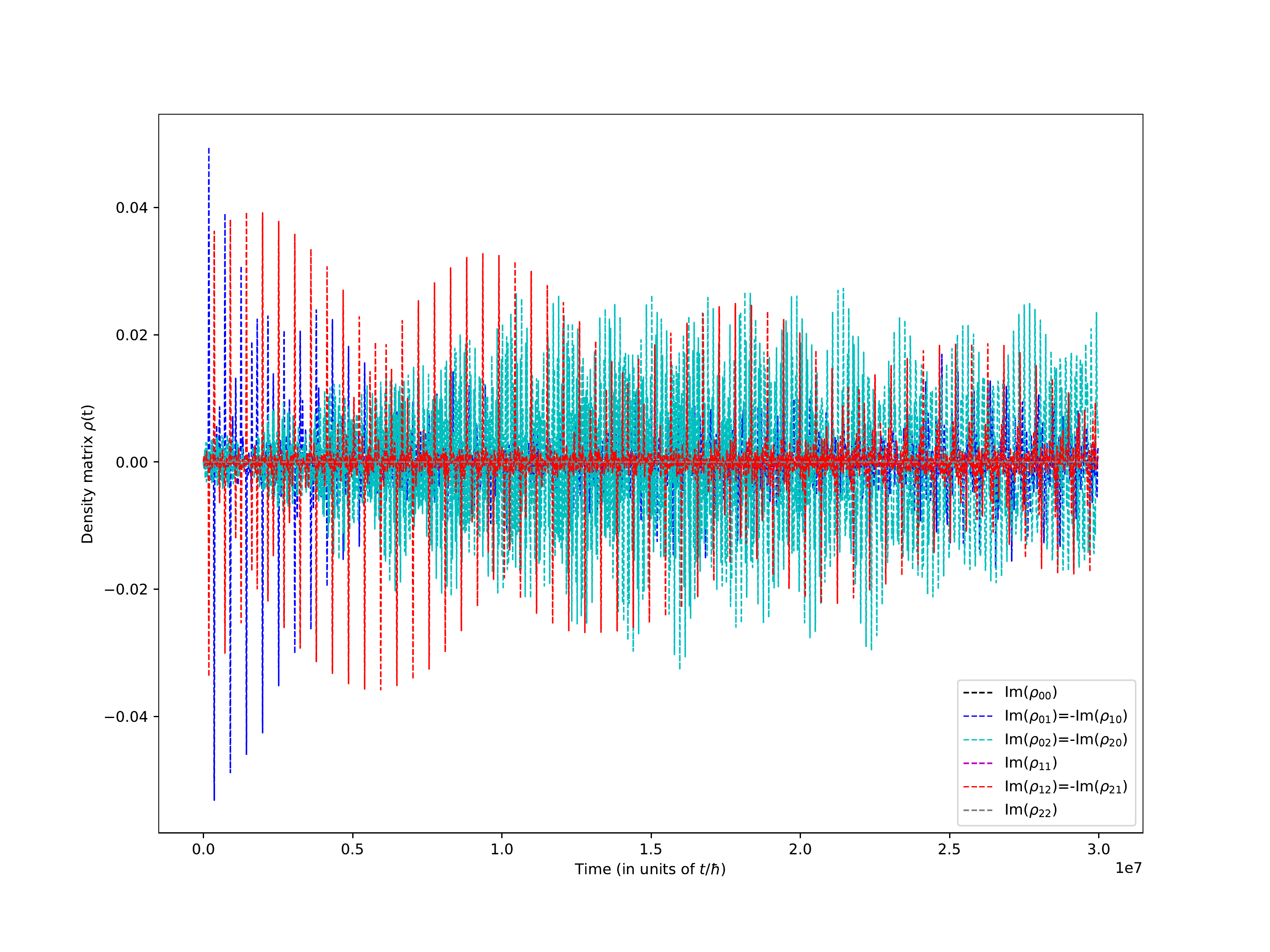}
\caption{Real (upper plot) and imaginary (lower plot) matrix elements (Eq.\ref{eq:rhoij}) of the density matrix as a function of time for the three-flavor scheme. We used the following parameters: $\sigma=20$ and $\lambda_{coup}=1.0$ for the Gaussian function and the coupling to the environment.  } 
\label{fig:rhot_3x3}
\end{figure}

\begin{figure}
\centering
\includegraphics[width=\hsize]{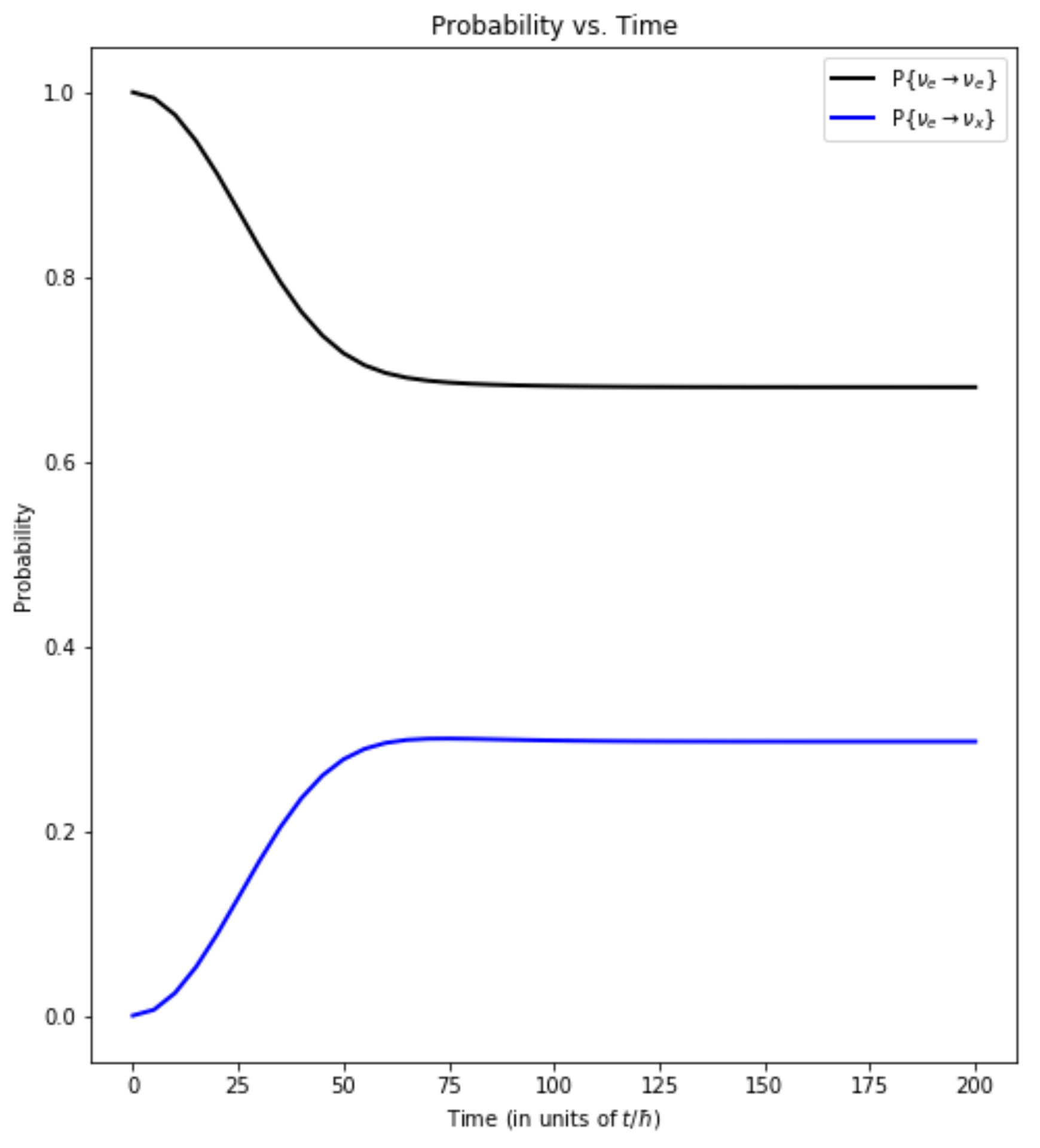}
\caption{Eigenvalues of the density matrix $\rho(t)$.From the initial values $P_{\nu_e \rightarrow \nu_e}(t=0)=1$ and $P_{\nu_e \rightarrow \nu_e}(t=0)=0$, the probabilities evolve to the asymptotic values $P_{\nu_e \rightarrow \nu_e}= 0.67871$ and $P_{\nu_e \rightarrow \nu_x}=0.32129$ of the pointer states.} 
\label{fig:prob3x3}
\end{figure}

Diagonalization of this matrix gives the survival probabilities:
\begin{equation}\label{Pe}
P_{\nu_e \rightarrow \nu_e}(t \rightarrow \infty)= 0.67871,\\
\end{equation}
\begin{equation}\label{Pmu}
P_{\nu_e \rightarrow \nu_{x}}(t \rightarrow \infty)= 0.32129.
\end{equation}
Fig. \ref{fig:rhot_3x3} shows the elements of the density matrix as a function of time, as the system evolves to the pointer states.The fact that the matrix $\rho(t)$ loses its pure-state nature due to decoherence is better illustrated by the results shown in Figure \ref{fig:prob3x3}, where the final pointer states are identified by means of the probabilities $P_{\nu_e \rightarrow \nu_e}$ and 
$P_{\nu_e \rightarrow \nu_{x}}$.

\begin{figure*}
\centering
\includegraphics[width=\hsize]{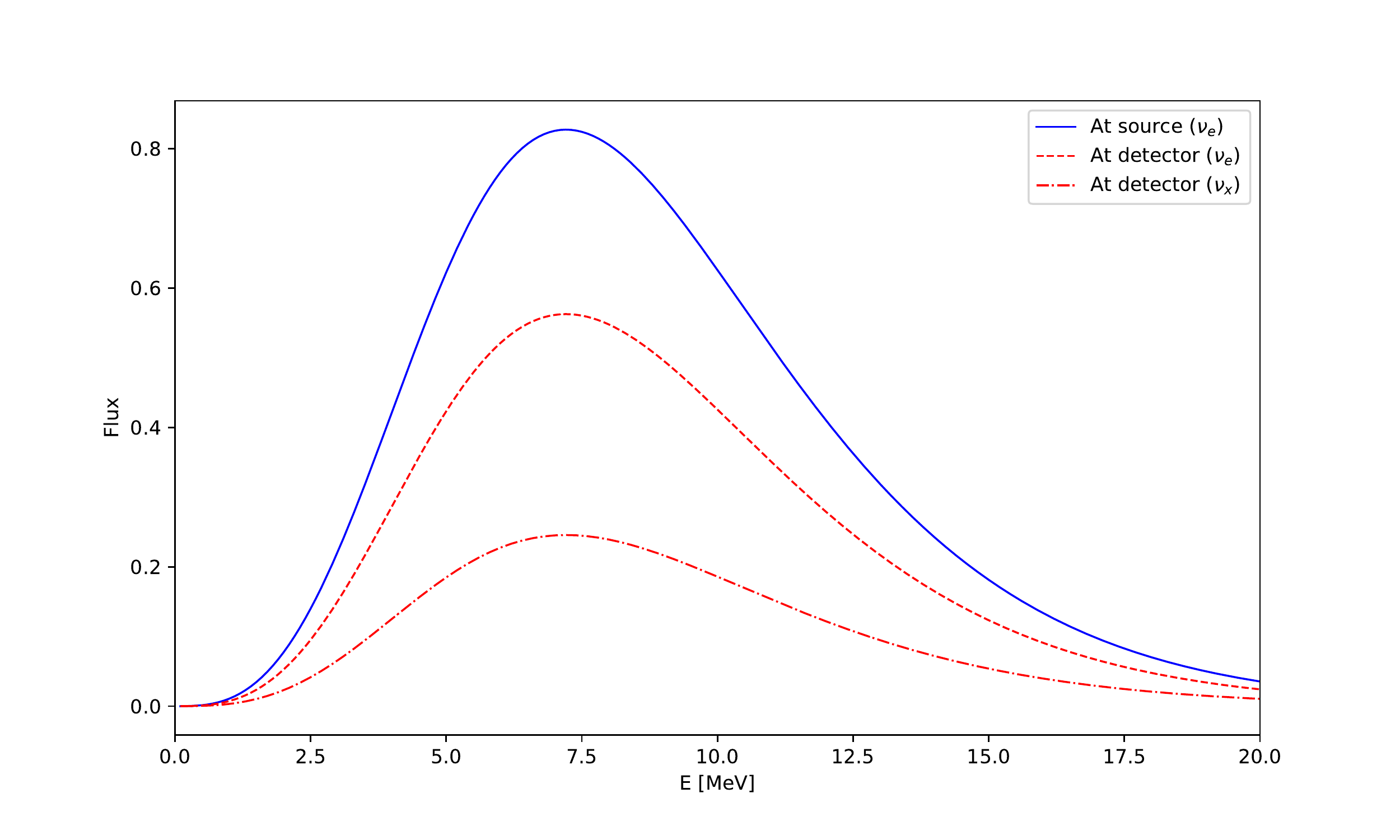}
\caption{Neutrino flux at the moment of the creation at source (blue, only electron neutrinos are created according to our model) and at the detector on Earth (red, due to decoherence effects, some electron neutrinos disappear and $x$ neutrinos are created).} 
\label{fig:flux_detector_3x3}
\end{figure*}

The neutrino flux, for neutrinos emitted at the source (see Fig.\ref{fig:flux}), should then be renormalised to account for the evolution of $\rho$ from pure to pointer states. This is done by multiplying the curves of Figure \ref{fig:flux} by the probabilities  (\ref{Pe}) and (\ref{Pmu}). The results are shown in Fig \ref{fig:flux_detector_3x3}.

We shall calculate the characteristic wavelengths of flavor-oscillations to compare them with the size of the regions where the effects take place.

In the case of flavor-oscillations, the amplitude of the electron-flavor survival $A_{\nu_{e} \rightarrow \nu_{e}}$ is given by
\begin{equation}
Re(A_{\nu_{e} \rightarrow \nu_{e}})= \abs{U_{\nu 1}}^2 + \sum_{j=2,3} \abs{U_{\nu j}}^2 cos(\Delta_{1j})
\end{equation}
and
\begin{equation}
Im(A_{\nu_{e} \rightarrow \nu_{e}})=\sum_{j=2,3} \abs{U_{\nu j}}^2 sin(\Delta_{1j}),
\end{equation}
for the real and imaginary parts of the amplitude, respectively, with 
\begin{equation}
\Delta_{1j}=\frac{(m_j^2-m_1^2)c^4 t}{2E\hbar}=\frac{\delta_{1j}^2 \, t}{2E\hbar}.
\end{equation}
To make a rough estimation of the period of oscillations, we take the squared mass difference $\delta_{12}^2$ in the normal hierarchy and write for the period $T_{osc}$ \citep{2016EPJC...76..339K}
\begin{equation}
T_{osc}=\frac{8 \pi  E  \hbar c}{\delta_{12}^2 \, c^4}.
\end{equation}
The corresponding wavelength for flavor oscillations will then be 
\begin{equation}
\lambda_{osc}=cT_{osc}.
\end{equation}
For the neutrino mean energy obtained in our calculations, $\langle E \rangle = 3.98$ MeV, we have $\lambda_{osc}=1.98 \times 10^8$ cm. This is much larger than the distance from the event horizon to the external part of the crust ($\approx 10^{5}$ cm), therefore confirming the pure-state nature of the density matrix (Eq.\ref{eq:rho0}) for neutrinos leaving the source.

\subsection{About the observability of the emitted neutrinos}

The results which we have presented so far show that the survival probabilities for the emitted electron-neutrinos change considerably, as do the calculated fluxes at the source and at the detector.
As mentioned before, Eq.\ref{eq:flux} gives the number of particles (electrons or neutrinos) with energies in the interval $E \pm \Delta E$.

\citet{2011JPhCS.309a2029K} performed a simulation of a Supernova event at a distance of 10 kpc with total emitted energy of $2.9 \times 10^{53}$ erg, starting from a $20$M$_\odot$ progenitor and considering inverse beta decay, neutron capture and positron annihilation as the main channels for neutrino interaction. They obtain a mean energy of the order of 15 MeV and a rate of $\approx 1.7 \times 10^5$ counts/s. In our case, we consider a NS-NS merger leading to a $2.7$M$_\odot$ progenitor at a redshift $z=0.9$, which corresponds to a distance of $\sim 3800$ Mpc (or $10^{28}$ cm, as stated in Table \ref{tab:parameters}), just like GRB 090510 \citep{2009GCN..9353....1R}. The only channel considered for neutrino production in our model is $e^--e^+$ annihilation (we intend to extend the model by considering more production channels that contribute to the total neutrino flux in a future work). We obtain a neutrino mean energy per particle of $3.98$ MeV. The rate at the detector is thus of the order of $10^{-4}$ events/s, which is far from being detected with the current Ice Cube sensitivity and effective area. However, this may be achieved by the future detector generations. What we would like to emphasize is that our calculations give us a mean neutrino energy which falls in the range of supernovae neutrinos (see Fig. 1 of \citet{2012EPJH...37..515S}).

\section{Conclusions}\label{conclusions}

In this work we have investigated the processes leading to the emission of neutrinos in short-GRB progenitors. Following the discussions advanced in the literature \citep{2008mgm..conf.1989B} we have modeled the system so that the $e^{-}e^{+}$-plasma is the main source of neutrinos. These neutrinos travel through the region between the BH event-horizon and the crust, their density matrix being that of pure states described by neutrino flavor-oscillations. Because of the astronomical scale of the distance between the source and the detector on Earth, decoherence effects due to interaction with the cosmic background may become important. We have calculated these effects by adopting a Gaussian model to incorporate the cosmic background. 

The present calculations give a mean neutrino energy which falls in the range of SN neutrinos. The value of the predicted neutrino flux is still far away from observation but considering the continuous advances in detector technology it could be reachable by future generations of experiments.

Further work is in progress concerning the time delay between neutrino and photon emission in GRBs.

\acknowledgments

The authors would like to thank Dr. A. Marinelli for useful discussions. This work has been partially supported by the National Research Council of Argentina (CONICET) by the grant PIP 616, and by the Agencia Nacional de Promoci\'on Cient\'ifica y Tecnol\'ogica (ANPCYT) PICT 140492. A.V.P and O.C. are members of the Scientific Research career of the CONICET.


\end{document}